\documentclass[conference]{IEEEtran}
\IEEEoverridecommandlockouts
\usepackage{cite}

\usepackage{threeparttable}
\usepackage{multirow}

\usepackage{url}
\usepackage{subfigure}
\usepackage{rotating}
\usepackage{soul}
\usepackage{amsmath,amssymb,amsfonts}
\usepackage{verbatim}

\usepackage[linesnumbered,ruled,vlined]{algorithm2e}
\usepackage{graphicx}
\usepackage{graphics}
\usepackage{textcomp}
\usepackage{xcolor}
\usepackage{utfsym}
\usepackage{makecell}
\def\BibTeX{{\rm B\kern-.05em{\sc i\kern-.025em b}\kern-.08em
    T\kern-.1667em\lower.7ex\hbox{E}\kern-.125emX}}

\usepackage{pifont}

\usepackage{upgreek}
\usepackage{textcomp}

\usepackage{enumitem}

\makeatletter

\newcommand{\Rmnum}[1]{\expandafter\@slowromancap\romannumeral #1@}
\makeatother

\usepackage{marvosym}
\begin{document}

%\title{Towards Efficient Dynamic Scheduling in Dataflow Architectures via Architecting the Task Flow Plane\\
%\title{StreamDCIM: Streaming Digital CIM Architecture for LLM Training through 3D Parallelism\\
    %\title{StreamDCIM: A Tile-based Streaming Digital Computing-in-Memory Accelerator for Sparse Transformer via Hardware-Software Co-design\\
    %\title{StreamDCIM: A Tile-based Streaming Digital CIM Accelerator for Multimodal Transformer via Hybrid Sparse Attention and Diagonal Sparse Compression \\
    \title{StreamDCIM: A Tile-based Streaming Digital CIM Accelerator with Mixed-stationary Cross-forwarding Dataflow for Multimodal Transformer\\
\vspace{-0.1cm}
}

\author{Shantian Qin, Ziqing Qiang, Zhihua Fan\textsuperscript{\Letter}, Wenming Li\textsuperscript{\Letter}, Xuejun An, Xiaochun Ye, Dongrui Fan\\
\textit{State Key Lab of Processors, Institute of Computing Technology, Chinese Academy of Sciences, Beijing, China}\\
\textit{School of Computer Science and Technology, University of Chinese Academy of Sciences, Beijing, China}\\
Email: \{qinshantian23s, qiangziqing23s, fanzhihua, liwenming, axj, yexiaochun, fandr\}@ict.ac.cn
\vspace{-0.3cm}
}

\maketitle

\begin{abstract}
%Dataflow architectures are considered promising architecture, offering a commendable balance of performance, efficiency, and flexibility. 
%Multimodal Transformers are emerging artificial intelligence (AI) models that comprehend a mixture of signals from different modalities like vision, natural language, and speech. 
%Digital computing-in-memory (CIM) architecture is considered a promising architecture with high efficiency and accuracy.
Multimodal Transformers are emerging artificial intelligence (AI) models designed to process a mixture of signals from diverse modalities. 
%Digital computing-in-memory (CIM) architecture is  as a potential hardware solution to reduce data movement with high accuracy.
Digital computing-in-memory (CIM) architectures are considered promising for achieving high efficiency while maintaining high accuracy. 
%are considered promising for achieving high efficiency and accuracy.
%However, current digital CIM-based accelerators face challenges due to inflexibility in microarchitecture, dataflow, and pipeline, limiting their ability to efficiently accelerate multimodal Transformers. 
However, current digital CIM-based accelerators exhibit inflexibility in microarchitecture, dataflow, and pipeline to effectively accelerate multimodal Transformer.
%to effectively accelerate multimodal Transformer. 
%In this paper, we propose GEMINI, a dataflow architecture with decoupled task flow and data flow planes, dedicated to efficient dynamic scheduling through hardware-software co-optimizations. 
%In this paper, we introduce GEMINI, a novel dataflow architecture that enhances dynamic scheduling through decoupled task flow and data flow planes, leveraging hardware-software co-optimizations.
In this paper, we propose StreamDCIM, a tile-based streaming digital CIM accelerator for multimodal Transformers. 
%StreamDIM introduces three levels of flexibility to overcome microarchitecture, dataflow, and pipeline challenges
It overcomes the above challenges with three features: 
First, we present a tile-based reconfigurable CIM macro microarchitecture with normal and hybrid reconfigurable modes to improve intra-macro CIM utilization. 
Second, we implement a mixed-stationary cross-forwarding dataflow with tile-based execution decoupling to exploit tile-level computation parallelism. 
Third, we introduce a ping-pong-like fine-grained compute-rewriting pipeline to overlap high-latency on-chip CIM rewriting. 
Experimental results show that StreamDCIM outperforms non-streaming and layer-based streaming CIM-based solutions by geomean 2.63$\times$ and 1.28$\times$ on typical multimodal Transformer models.
%We implement PANDA in RTL design and %demonstrate 
%Experimental results show that in a wide range of real-world applications,
%including scientific computing, artificial intelligence, digital signal processing, and graph processing algorithms, 
%GEMINI achieves up to 1.89$\times$ performance improvement and 1.67$\times$ energy efficiency improvement over the state-of-the-art dataflow architectures.
%compared to the state-of-the-art architectures, PANDA outperforms REVEL, Plasticine, DFU, and MTDE by geomean 2.53$\times$, 1.90$\times$, 1.38$\times$, and 1.19$\times$.
\end{abstract}


\vspace{-0.15cm}
\section{Introduction}
\vspace{-0.05cm}

Transformers, a type of neural network (NN) model, have achieved remarkable success across a wide range of artificial intelligence (AI) tasks, outperforming recurrent  neural networks (RNNs) and traditional convolution neural networks (CNNs) in both natural language processing (NLP) \cite{NLP} and computer vision (CV) \cite{CV}. Their exceptional performance is largely due to the attention mechanism, which effectively capture contextual knowledge from the entire input sequence. 
Additionally, a key goal in AI is to emulate human multimodal perception, enabling systems to comprehend information across various modalities, such as language and vision. Recent algorithmic advancements have highlighted the potential of multimodal Transformers in learning from diverse inputs, delivering impressive results in tasks like multilingual image retrieval and action prediction \cite{Multimodal21,Multimodal22}.

\begin{figure}[h]
\vspace{0.2cm}
\centering
  \includegraphics[width=0.485\textwidth]{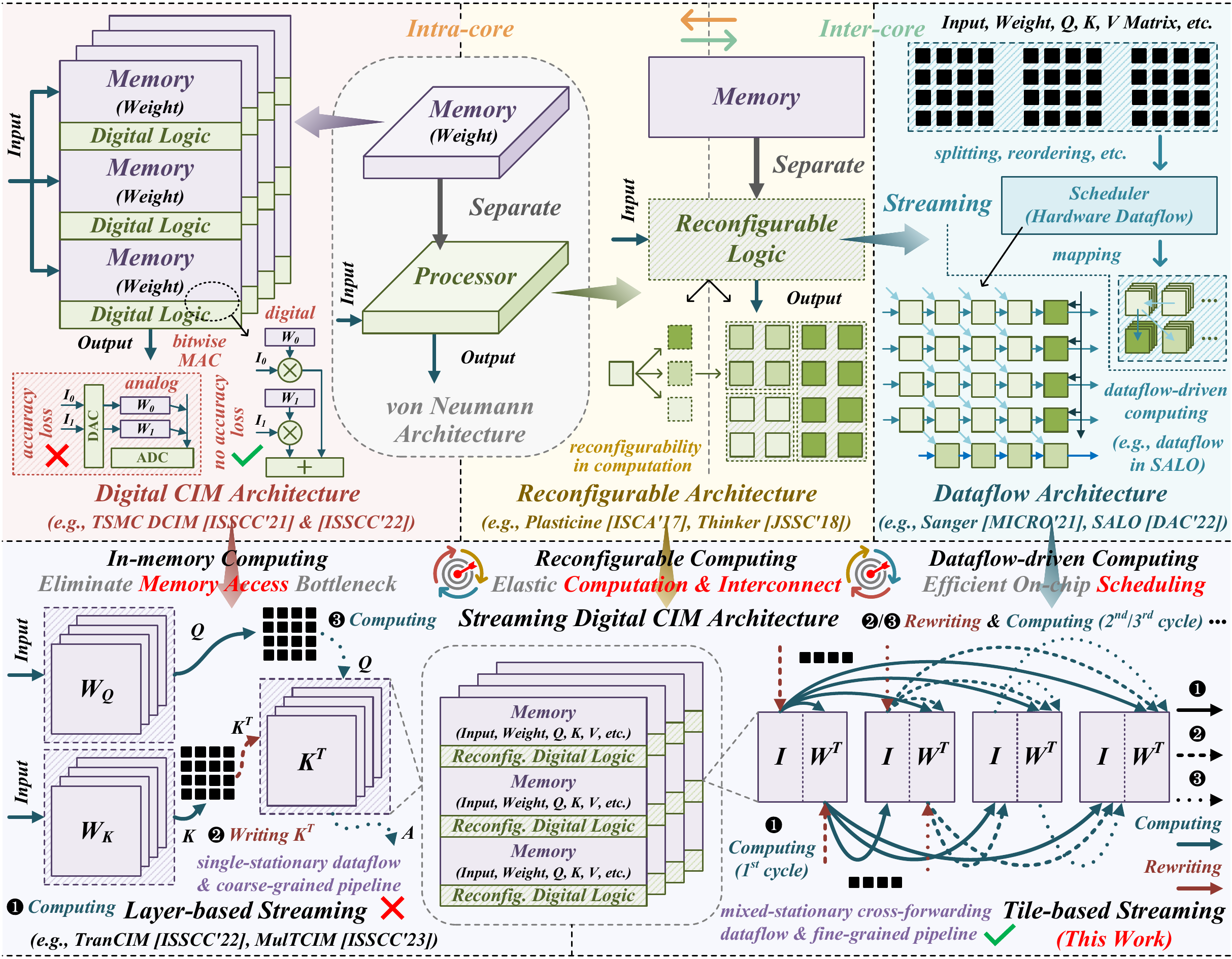}
  \setlength{\abovecaptionskip}{-0.35cm}
  \setlength{\belowcaptionskip}{-0.35cm}
  \caption{Overview of Different NN Accelerator Architectures.}
  \label{abstract}
  \vspace{-0.6cm}
\end{figure}

Abundant prior works have been proposed to accelerate NNs through architectural optimizations. Fig. \ref{abstract} provides an overview of different NN accelerator architectures. Conventional von Neumann architectures have separate computation and memory units. Some works eliminate the memory access bottleneck by integrating multiply-and-accumulate operations directly into memory \cite{TSMC21,TSMC22}, while others leverage reconfigurable digital logic to enable flexible computation \cite{Plasticine,Thinker}. Additionally, several solutions enhance the efficiency of on-chip scheduling for both computation and memory access through well-designed hardware dataflow \cite{Sanger,SALO}.

\begin{figure*}[h]
\centering
  \includegraphics[width=\textwidth]{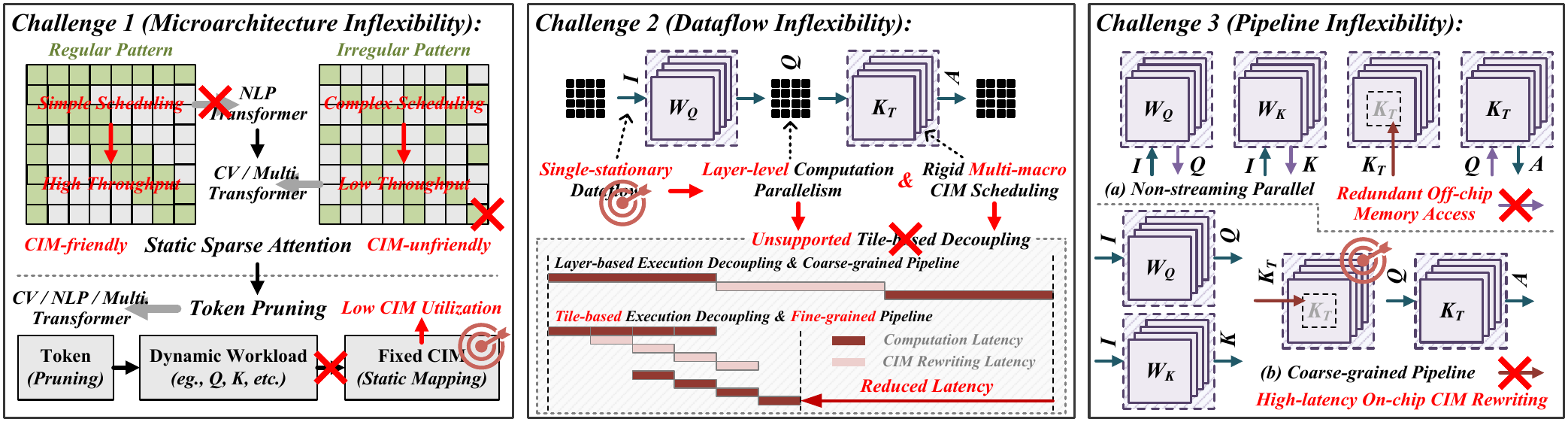}
  \setlength{\abovecaptionskip}{-0.2cm}
  \setlength{\belowcaptionskip}{-0.2cm}
  \caption{Challenges for CIM-based Multimodal Transformer Acceleration: 1) Microarchitecture Inflexibility. 2) Dataflow Inflexibility. 3) Pipeline Inflexibility.}
  \label{challenges}
  \vspace{-0.6cm}
\end{figure*}

Digital computing-in-memory (CIM) architecture combines the advantages of both CIM and digital architectures, achieving high efficiency by embedding computation directly within memory while ensuring high accuracy by eliminating analog non-ideality through its digital in-memory logic \cite{ReDCIM,TranCIM,MulTCIM,SparseTrans}. Furthermore, unlike analog CIM, digital CIM offers greater customization, allowing for optimization of the in-memory logic. This flexibility makes digital CIM a promising solution to integrate the benefits of in-memory computing, reconfigurable computing, and dataflow-driven computing paradigms, ultimately leading to a streaming digital CIM architecture. Such an architecture holds promise to accommodate efficient memory access, computation and scheduling. The pipeline and parallel reconfigurable modes in TranCIM \cite{TranCIM} illustrate layer-based streaming, enhancing overall performance.

However, the multimodal Transformer introduces new challenges for existing digital CIM-based accelerators.
%, as it also has substantial MM computation demands and introduces a cross-modal attention mechanism to vanilla Transformers that enables joint information learning from various modalities. 
% (?不确定要不要加)哪怕是上述的layer-based streaming也受限于low hardware utilization due to single-sationary dataflow和on-chip writing latency due to coarse-grained pipeline。
Our observations indicate that current digital CIM-based accelerators exhibit \textit{\textbf{inflexibility in microarchitecture, dataflow, and pipeline}} to effectively accelerate multimodal Transformer (see Fig. \ref{challenges}).%, as illustrated in Fig. \ref{challenges}.
\textit{Challenge 1 (Microarchitecture Inflexibility):} Previous digital CIM-based Transformer accelerators mainly rely on static sparse attention, overlooking the benefits of dynamic token pruning, or they lack the flexible microarchitecture for efficient token pruning. 
Static sparse attention limits attention computations to a predefined set of token pairs, reducing the computational and memory complexity in Transformer models, and is widely used in CIM-based Transformer accelerators \cite{TranCIM,MulTCIM,SparseTrans}.
This approach typically employs a fixed attention pattern for a given task, as it is impossible to predetermine which tokens should be attended to each other \cite{Longformer,BigBird,ETC}. 
While static sparse attention is well-suited for NLP applications with long token sequences \cite{LongSparse}, 
%For CV tasks, the image is divided into independent patches, and then mapped as token sequences in order. 
%These tokens have special spatial characteristics corresponding to the image itself, making the static sparse attention more complicated \cite{Evo-ViT,DynamicViT}. 
%This results in lower throughput in vision transformer even with harder scheduling, especially for CIM-based accelerator normally with rigid weight-stationary dataflow. 
it often results in irregular patterns in vision and multimodal Transformers due to the complex spatial relationships in images and dynamic token interactions across different modalities. 
These irregularities hinder achieving high throughput, even with complex scheduling method, particularly for CIM-based Transformer accelerators using rigid weight-stationary dataflows \cite{DynamicViT,Evo-ViT}.
In contrast, dynamic token pruning offers a more adaptive solution, optimizing tasks in NLP and CV \cite{SpAtten,Evo-ViT}.
%In short, these limitations create a difficult tradeoff between throughput and accuracy in previous works, and render them unsuitable for vision Transformers.
Multimodal inputs, such as language and images, contain tokens of varying significance \cite{SpAtten,DynamicViT,Evo-ViT}. 
%token pruning is a dynamic sparsity technology unique to Transformer models that is suitable for optimizing NLP and CV tasks \cite{SpAtten,Evo-ViT,DynamicViT}. 
%Plenty of unessential tokens exist in human languages, which can be pruned away to boost efficiency. Many image patches contribute very little to the final prediction in vision Transformers. 
Token pruning selectively retains attentive tokens and gradually prune inattentive ones across layers. 
Experimental results show that pruning the redundancy in image tokens can lead to over 1.6$\times$ speedup with negligible accuracy loss \cite{Evo-ViT}.
%layer-based inter-macro reconfigurable architecture
%从token pruning会对QK矩阵的规模会在运行时动态变化，导致之前的固定fixed intra-macro CIM microarchitecture 和 compute mapping strategy 会导致CIM的资源利用率低
However, token pruning introduces dynamic workloads like \textit{Matrix} $Q$, $V$, $K$ generation during runtime. In this context, the fixed intra-macro CIM microarchitecture and static workload mapping can result in low CIM utilization.
% 视觉transformer和多模态transformer的静态稀疏注意力pattern往往是irregular的，对于这种pattern即使通过复杂调度，其吞吐量依旧不高

\textit{Challenge 2 (Dataflow Inflexibility):} 
Previous digital CIM-based Transformer accelerators often employ fixed single-stationary dataflows, such as weight-stationary dataflow \cite{TranCIM,MulTCIM}, which are sub-optimal \cite{MobiLattice,MorphableCIM} and limited by layer-level computation parallelism and rigid multi-macro CIM scheduling. This limitation restricts support for tile-based execution decoupling and tile-level computation parallelism, ultimately hindering overall performance.
% fixed single-stationary dataflow
% Previous CIM-based Transformer accelerators have primarily relied on static sparse attention, overlooking the benefits of dynamic token pruning, or they lack the flexible microarchitecture for efficient token pruning. 

%Challenge 3 (Pipeline Inflexibility): Attention layers introduce dynamic matrix multiplications (QKT and PV), where both weights and inputs are generated at runtime. This leads to redundant off-chip memory access for intermediate data in traditional CIM-based Transformer accelerators using non-streaming parallel methods [5], [24], [25]. While TranCIM [12] reduces redundant off-chip accesses through a coarse-grained pipeline, it still faces significant pipeline bubbles and incurs additional on-chip CIM rewriting, resulting in high latency and energy consumption. For example, the benchmark involves QKT with INT8 precision and a K matrix size of 2048×512. With a 512-bit memory access bandwidth, TranCIM experiences over 57% latency in rewriting the K matrix in CIM macros during QKT computation. Furthermore, considering the generation of Q and K, QKT accounts for 66.7% of computations, with CIM rewriting contributing to 88.9% of the latency.

\textit{Challenge 3 (Pipeline Inflexibility):} 
Attention layers introduce dynamic matrix multiplications (such as $QK^{T}$), where both weights and inputs are generated at runtime. This leads to redundant off-chip memory access for intermediate data in traditional CIM-based Transformer accelerators using non-streaming parallel methods \cite{TSMC21,Non-streaming-2,Non-streaming-3}. 
%TranCIM \cite{TranCIM} reduces the redundant off-chip memory access through the coarse-grained pipeline but typically exists a significant number of pipeline bubbles and introduces additional on-chip CIM rewriting with high latency and energy consumption.
Although TranCIM \cite{TranCIM} mitigates redundant off-chip accesses through pipeline and parallel reconfigurable modes, it still encounters significant pipeline bubbles and incurs additional on-chip CIM rewriting due to its coarse-grained pipeline, resulting in high latency and energy consumption. 
For example, assuming a 512-bit memory access bandwidth and a benchmark involving $QK^{T}$ with INT8 precision and a $K$ matrix size of 2048$\times$512, TranCIM incurs over 57\% latency to rewrite the $K$ matrix in CIM macros during $QK^{T}$computation. When considering $Q$ and $K$ generation, $QK^{T}$comprises 66.7\% of computations, with CIM rewriting accounting for 88.9\% of the latency \cite{CIMRing}.

In this paper, we propose StreamDCIM, a tile-based streaming digital CIM accelerator for multimodal Transformers. 
%It overcomes the above challenges with three key features.
StreamDCIM introduces three levels of flexibility to overcome the above challenges.
%StreamDIM introduces three levels of flexibility to overcome challenges in microarchitecture, dataflow, and pipeline: \textit{1)} a tile-based reconfigurable CIM macro \textbf{microarchitecture} with normal and hybrid reconfigurable modes to improve intra-macro utilization, \textit{2)} a mixed-stationary cross-forwarding \textbf{dataflow} to exploit tile-level computation parallelism, and \textit{3)} a ping-pong-like fine-grained compute-rewriting \textbf{pipeline} to overlap high-latency CIM rewriting.

\begin{enumerate}[label=\arabic*)]
\setlength{\itemindent}{\parindent}
% 这一部分明天再看看之前和强姐怎么写的

%We describe an ISA and its programmability, supporting the prefetch/postback and load/store for prefetchable and non-prefetchable data. It fosters the decoupling of the prefetchable data access from the application execution.
\item For \textbf{Challenge 1}, we present a tile-based reconfigurable CIM (TBR-CIM) macro \textbf{\textit{microarchitecture}} with normal and hrbrid reconfigurable modes to improve the intra-macro CIM utilization.
% 计算并行性不够 - 解决方案：从同时算一行到同时算一行加一列
% 提出了 Intra-macro reconfigurable Digital CIM Marco Arichitecture，支持tile-based的Ring状streaming与computing，可以更细粒度地充分利用计算资源

%We introduce an on-chip memory microarchitecture tailored for multi-core accelerators based on the ROMA ISA. It leverages reconfigurable physical storage shared by SPM and cache to reduce on-chip area overhead.
\item For \textbf{Challenge 2}, we implement a mixed-statinary cross-forwarding \textbf{\textit{dataflow}} with tile-based execution decoupling and elastic single-macro scheduling to exploit the tile-level computation parallelism.
%enhancing tile-based data reuse efficiency and computation parallelism.
% reloading时间很长 - 解决方案，用计算时间掩盖reloading时间
% 提出了 mixed-stationary（感觉这里不太合适？）的dataflow，既减少了一部分的reloading（不确定？），同时也通过设计高效的computing-reloading流水线对不可避免的reloading进行了overlapping

%\item We propose an adaptive semi-centralized dynamic task scheduling policy for GEMINI architecture.
%\item We propose an adaptive load balancing scheme that integrates dataflow-driven dynamic issuing and throughput-aware dynamic balancing mechanisms.
\item For \textbf{Challenge 3}, we introduce a ping-pong-like fine-grained compute-rewriting \textbf{\textit{pipeline}} to overlap the high latency of on-chip CIM rewriting.
% 多模态的输入长度不同 - 解决问题，对于cross-mode的SA计算，两个一起算，就不用区分X和Y了
% 提出了一个调度器，用于协调单模态和多模态的self-attention计算的输入和调度

%\item (Overall Workflow) 
%We present a mechanism for data mapping and on-chip memory partitioning in applications targeting ROMA-based architectures. It facilitates the organized separation of data into prefetchable and non-prefetchable categories.

%\item We implement all modules of StreamDCIM using Verilog. 
%Experimental results show that StreamDCIM outperforms state-of-the-art dataflow architectures, including Plasticine, ParallelXL, and MTDE, by geomean 1.89$\times$, 1.47$\times$, and 1.21$\times$ in a wide range of real-world applications.
%Experimental results show that SALO achieves 17.66x and 89.33x speedup on average compared to GPU and CPU implementations, respectively, on typical workloads like Longformer and ViL.

\end{enumerate}

%We implement all modules of StreamDCIM using Verilog. 
Experimental results show that StreamDCIM outperforms non-streaming and layer-based streaming CIM-based solutions by geomean 2.63$\times$ and 1.28$\times$ on typical multimodal Transformer models. 
%It reduces fine-tuning energy by 4.27× and offers 3.57× speedup for GPT-2. 
%Experimental results show that StreamDCIM outperforms state-of-the-art dataflow architectures, including Plasticine, ParallelXL, and MTDE, by geomean 1.89$\times$, 1.47$\times$, and 1.21$\times$ in a wide range of real-world applications.

\vspace{-0.05cm}
\section{StreamDCIM Designs}
\vspace{-0.05cm}

\begin{figure*}[h]
\centering
  \includegraphics[width=0.97\textwidth]{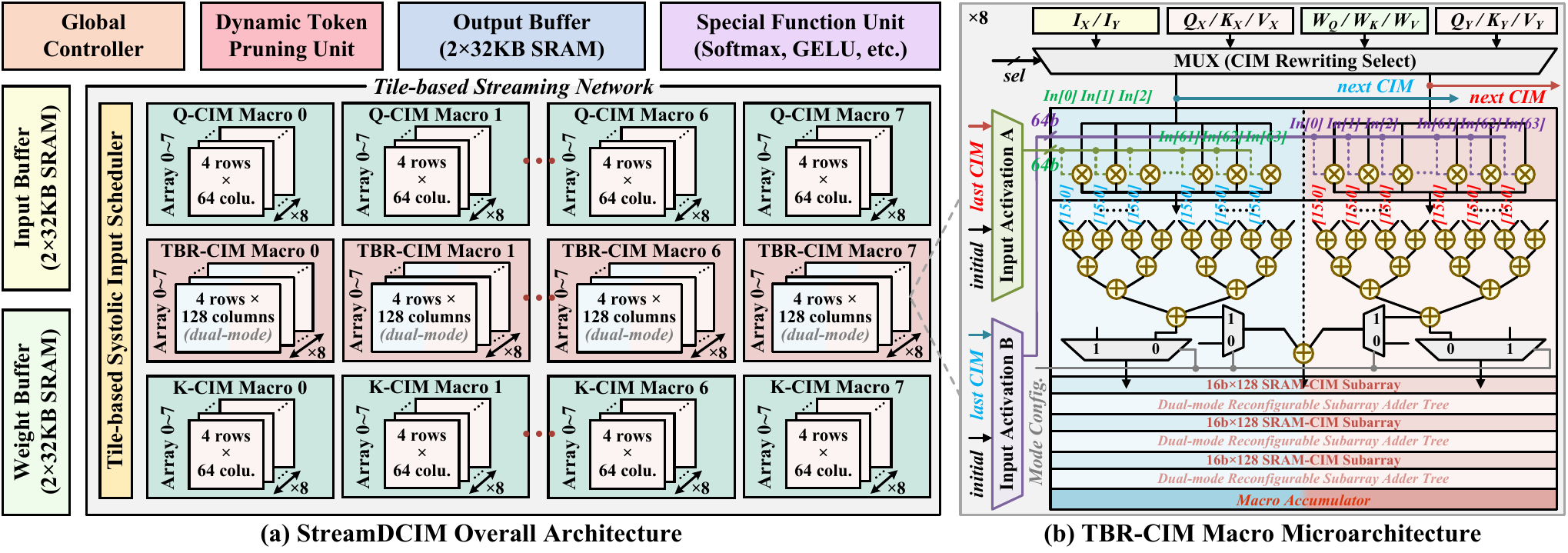}
  \setlength{\abovecaptionskip}{-0.05cm}
  \setlength{\belowcaptionskip}{-0.05cm}
  \caption{StreamDCIM: (a) Overall Architecture. (b) TBR-CIM Macro Microarchitecture.}
  \label{StreamDCIM}
  \vspace{-0.6cm}
\end{figure*}

\textbf{\textit{Multimodal Transformer \& Attention Mechanism:}} 
% attention mechanism
The input to an attention layer in vanilla Transformers is typically a sequence of $N$ tokens ($I$). 
By multiplying $I$ with weight matrices $W_{Q}$, $W_{K}$, $W_{V}$, we obtain the query ($Q$), key ($K$), and value ($V$) matrices. 
Next, the attention matrix ($A$) is obtained by multiplying $Q$ and the transpose of $K$ ($K_{T}$). $A$ is then normalized using the softmax function to yield a probability matrix ($P$), which is finally multiplied by $V$ to generate the output. 
% multimodal transformer
%In addition, a typical multimodal Transformer model has a stacking structure of single-modal and cross-modal encoders. 
In multimodal Transformers, the structure often consists of stacked single-modal and cross-modal encoders to process different modalities, such as vision and language. To process two modalities, the encoders can be divided into two streams for modal $X$ and modal $Y$. 
%For example, $X$ is for vision, and $Y$ is for language. 
Single-modal attention layers are similar to those in vanilla Transformers. 
Cross-modal attention layers are introduced to facilitate information exchange between the two modalities. 
For instance, in the stream for Modal $X$, 
$Q_{X}$ come from modal $X$ ($I_{X}W_{Q}$), while $K_{Y}$ and $V_{Y}$ are from modal $Y$ ($I_{Y}W_{K}$ and $I_{Y}W_{V}$). 
A similar process occurs for modal $Y$.
The overall architecture of the StreamDCIM accelerator is illustrated in Fig. \ref{StreamDCIM} (a). It comprises a tile-based streaming network (TBSN) with a tile-based systolic input scheduler and three CIM cores: Q-CIM, K-CIM, and TBR-CIM. Additionally, it includes a 64-KB input buffer, a 64-KB weight buffer, a 64-KB output buffer, a dynamic token pruning unit (DTPU), a special function unit (SFU), and a global controller. The CIM cores are interconnected through the TBSN’s pipeline bus, with each core containing eight CIM macros.% (TBR-CIM array size: 4$\times$16b$\times$128, Q/K-CIM array size: 4$\times$16b$\times$64). 

\vspace{-0.05cm}
\subsection{Tile-based Reconfigurable CIM Macro Microarchitecture}
\vspace{-0.05cm}

\begin{figure}[h]
%\vspace{0.2cm}
\centering
  \includegraphics[width=0.45\textwidth]{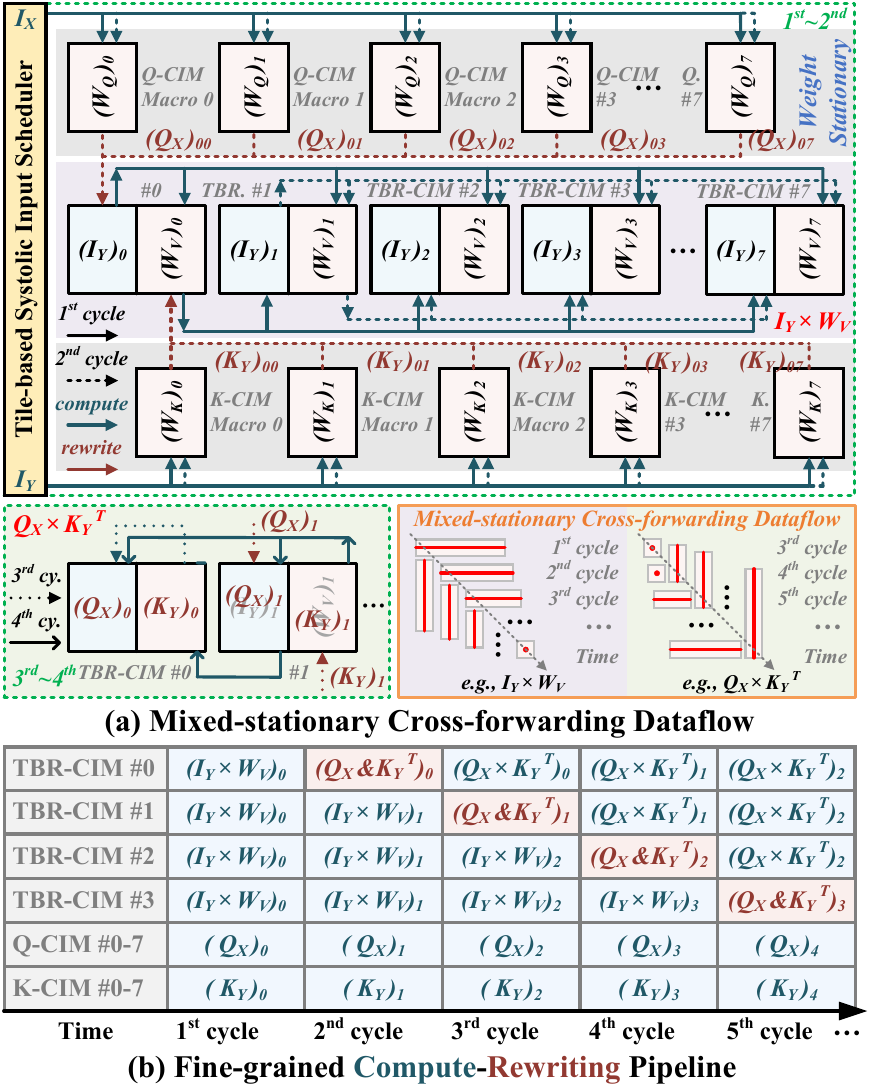}
  \setlength{\abovecaptionskip}{-0.15cm}
  \setlength{\belowcaptionskip}{-0.15cm}
  \caption{(a) Mixed-stationary Cross-forwarding Dataflow \& (b) Fine-grained Compute-Rewriting Pipeline (Example of the Stream for Modal X).}
  \label{Dataflow}
  \vspace{-0.7cm}
\end{figure}

Token pruning reduces computational complexity by eliminating redundant token. 
This process generally includes token ranking and selection, which can leverage attention probabilities, as the probability can indicate a token's relevance to all other tokens \cite{SpAtten,Evo-ViT}.
Each token’s importance rank can be computed by taking the column mean of its attention probability. 
Once a token is pruned, its $Q$, $K$, $V$ will be excluded from further computations. 

\begin{comment}
The TPAR aims to determine the most efficient approach for generating the Q matrix from different reformulation techniques, taking into account the token-pruning ratio. 
Similar to Evo-ViT\cite{Evo-ViT} and SpAtten\cite{SpAtten}, StreamDCIM uses the attention possibility as a natural indicator for identifying the significance of tokens in vision and language Transformers, removing redundant tokens. 
We observe that more redundant tokens can be safely removed in deeper layers, and fewer in shallower layers. It implies that the MHSA aggregates different tokens layer by layer, and a large number of similar tokens are produced in the process. 
While progressively pruning tokens, the CIM macros that previously store X can be reconfigured as WQ-CIM. In this configuration, the total memory capacity of WQ-CIM varies with the pruning ratio, which requires different types of reformulation.
\end{comment}

%Fig. 11 demonstrates the hierarchical and reconfigurable in-memory accumulators in a BM2-CIM macro. The 32 × 48 macro is divided into four 8 × 48 subarrays, so each macro has four rows of subarray adders and one macro accumulator. The minimum weight recision in ReDCIM is 8 b, so the macro’s 48 columns are divided into six groups

To facilitate efficient dynamic token pruning while maintaining high CIM utilization, the DTPU and TBR-CIM macro with normal and hybird reconfigurable modes are proposed, as illustrated in Fig. \ref{StreamDCIM}. %The TBR-CIM macro consists of three key modules: the \textit{Task Management Unit}, the \textit{Memory Unit}, and the \textit{Function Unit}.
% hybrid模式适合在token较多的前几层，normal 模式适合多次pruning之后的token较少时。这种可重构可以更好地适配动态token剪枝带来的workload减少at runtime，提高CIM的资源利用率。
Similar to Evo-ViT\cite{Evo-ViT} and SpAtten\cite{SpAtten}, StreamDCIM uses the attention possibility to identify the significance of tokens in various modalities, removing redundant tokens under the control of DTPU.
Each TBR-CIM macro consists of eight 4$\times$16b$\times$128 SRAM-CIM arrays and each array has four rows of dual-mode reconfigurable subarray adder trees and one macro accumulator. 
%We observe that more redundant tokens can be safely removed in deeper layers, and fewer in shallower layers. 
%It implies that the MHSA aggregates different tokens layer by layer, and a large number of similar tokens are produced in the process. 
%While progressively pruning tokens, the CIM macros that previously store $I$ and $W$ with the hybrid mode can be reconfigured as the CIM macros that only store $W$ with the normal mode. %In this configuration, the total memory capacity of WQ-CIM varies with the pruning ratio, which requires different types of reformulation.
As token pruning progresses, TBR-CIM macros that initially store both $I$ and $W$ in hybrid mode ($mode\_config = 0$) can be reconfigured to operate in normal mode ($mode\_config = 1$), storing only $W$. 
In normal mode, the TBR-CIM macro functions as a weight-stationary CIM macro, accelerating the $Q$, $K$, $V$ generation.

\vspace{-0.05cm}
\subsection{Mixed-stationary Cross-forwarding Dataflow}
\vspace{-0.05cm}

\begin{figure}%[htbp]
  \vspace{-0.3cm}
\centering
  \includegraphics[width=0.43\textwidth]{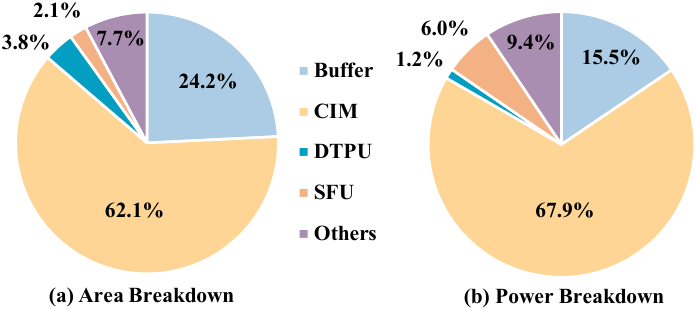}
  \vspace{-0.2cm}
  \caption{StreamDCIM: (a) Area Breakdown. (b) Power Breakdown.}
  \label{A&P}
  \vspace{-0.6cm}
\end{figure}

To support the tile-based execution decoupling, the mixed-stationary cross-forwarding dataflow is introduced, as shown in Fig. \ref{Dataflow} (a) for an example of modal $X$. 
%This dataflow strategy manages data movement efficiently by enabling simultaneous row-wise and column-wise data propagation across CIM macros. 
For instance, in the $1^{st}$ cycle, TBR-CIM $\#0$ provides row-wise $I_Y$ and column-wise $W_V$ as inputs to other CIM macros.  
Each row from $(I_Y)_{0}$ is sent to the $W_V$ part of TBR-CIM $\#0$-$7$, generating a full row in $V$. 
Concurrently, each column from $(W_V)_{0}$ is sent to the $I_Y$ part of TBR-CIM $\#1$-$7$, producing a portion of a column in $V$. 
%This cross-forwarding approach leverages the mixed-stationary configuration, where data is kept stationary in optimal segments to facilitate efficient, parallel updates across the CIM array, maximizing throughput in the tile-based execution model. 
This cross-forwarding computation approach leverages the mixed-stationary TBR-CIM configuration, facilitating more frequent reuse of stored data and utilizing tile-level computation parallelism. 
%facilitating tile-based execution decoupling and utilizing tile-level computation parallelism. 
Additionally, the calculation of $Q_{X}K_{Y}^{T}$ operates as the inverse of the process for $I_{Y}W_{V}$, which also exploits the mixed-stationary cross-forwarding dataflow. Conversely, the calculation of $I_{X}W_{Q}$ and $I_{Y}W_{K}$ follows the fixed weight-stationary dataflow.

%Weight-stationary $Q$ and $K$ generation

%Mixed-stationary cross-forwarding dataflow: $I_{Y}$$\times$$W_{V}$ and $Q_{X}$$\times$$K_{Y}^{T}$ 

%\subsection{Overall Workflow}
\subsection{Ping-pong-like Fine-grained Compute-Rewriting Pipeline}

To overlap the high latency associated with on-chip CIM rewriting, the ping-pong-like fine-grained compute-rewriting pipeline is proposed, as shown in Fig. \ref{Dataflow} (b). 
%In Pipeline StageD, the CIM macros have to hold infrequently used weights for a long time, leading to rigid multi-macro CIM scheduling.
%Thus, inputs and weights stored in CIM can be reused more frequently to improve CIM utilization. 
The above mixed-stationary cross-forwarding dataflow enables tile-based computation parallelism and elastic single-macro CIM scheduling, providing the possibility for a more finer-grained pipeline. 
In this pipeline, specific resources can be dynamically released and reallocated. 
For instance, in the $2^{nd}$ cycle, the inputs and weights stored in TBR-CIM $\#0$ are no longer required for ongoing calculations, freeing TBR-CIM $\#0$ for immediate rewriting. 
Meanwhile, $(Q_X)_0$ and $(K_Y)_0$ computations have been completed in Q-CIM $\#0$-$7$ and K-CIM $\#0$-$7$, respectively. 
%In this case, the rewriting of $(Q_X)_0$ and $(K_Y)_0$ and the computation of $(Q_X)_1$, $(K_Y)_1$, and $(V_Y)_1$ can be executed in parallel, overlapping the rewriting latency and enhancing the overall throughput. 
This timing permits the parallel execution of rewriting $(Q_X)_0$ and $(K_Y)_0$ alongside new computations of $(Q_X)_1$, $(K_Y)_1$, and $(V_Y)_1$.
Such parallelism effectively overlaps rewriting latency, enhancing throughput by ensuring continuous resource utilization across cycles.
%讲下QKV生成和QKT计算的掩盖，以及QK生成的rewriting

%Ping-pong-like Rewriting Overlapping:

%The SFU performs scaling and softmax on A to generate Matrix A = softmax((A/dk)). The A V-pipeline has a similar dataflow, except that the 2nd stage’s input A is loaded from the global buffer instead of the 1st stage. SEngines are combined to compute Matrix V, while DEngine computes the attention layer’s output matrix Att = AV. In this article, we will mainly use the QKT MMas example to explain the technical features of TranCIM.

\section{Experimental Results}

\subsection{Experiment Settings}

\textbf{Methodology.} 
%We implement ROMA microarchitecture in Verilog. We synthesize the design with Synopsys Design Compiler with a commercial 16nm technology with a 1GHz frequency. We use Synopsys PrimeTime PX for accurate power analysis. Table \ref{setup} shows our baseline architecture. For a detailed justification of the design choices see\cite{LRP}. 
%We develop a cycle-accurate micro-architecture simulator for hardware utilization and performance evaluation. The simulator is developed in C language based on SimICT framework [16] and can simulate behaviors such as memory access, data transfer, scheduling, etc. We calibrate the error to within ±7\% using RTL environment. We also implement our architecture using Verilog. We use Synopsys Design Compiler and a TSMC 28nm GP standard VT library to synthesize it and obtain area, delay and energy consumption, which meets timing at 1GHz. Table \ref{setup} shows the hardware parameters.
We implement all modules of StreamDCIM using Verilog. We synthesize the design using Synopsys Design Compiler with a commercial 28nm technology, meeting timing at a 200 MHz frequency. We use Synopsys PrimeTime PX for accurate power analysis. %Table \ref{hardware} shows the hardware design parameters, as well as the evaluated area and power breakdown of essential modules in StreamDCIM. %The DFGC simulator was also developed and was calibrated to cycle-level accuracy.
%Evaluation shows that GEMINI has an area footprint of 2.249 mm$^2$ in a 28nm process, and consumes a maximum power of 1.491 W.
%\textbf{Benchmark.} 
To evaluate our design, we select two typical multimodal Transformer models, ViLBERT-base and ViLBERT-large \cite{ViLBERT}, with the visual question answering (VQA) v2.0 dataset. To maintain high accuracy, the attention layers use INT16 precision.  In our configuration, \textit{X} is the vision modality, while \textit{Y} is the language modality, each with a token count of $N_X = N_Y = 4096$.

\textbf{Comparison.} %Our comparison is between semi-centralized dynamic scheduling (GEMINI), static scheduling (Plasticine \cite{Plasticine}), centralized dynamic scheduling (MTDE \cite{MTDE}), and decentralized dynamic scheduling (work-stealing). For work-stealing, our implementation is based on the description in ParallelXL \cite{Plasticine}.
We compare StreamDCIM using tile-based streaming solution (\textbf{Tile-stream}) with two digital CIM-based accelerators: non-streaming solution and layer-based streaming solution. 
The non-streaming solution (\textbf{Non-stream}) operates similarly to the conventional work mode of previous CIM accelerators \cite{TSMC21,Non-streaming-2,Non-streaming-3}. 
In Non-stream, dynamic matrix multiplication in the attention layers lead to redundant off-chip memory access for intermediate data, which negatively impacts performance and efficiency. 
The layer-based streaming solution (\textbf{Layer-stream}) aligns with the parallel and pipeline reconfigurable mode of TranCIM \cite{TranCIM}. 
In Layer-stream, on-chip CIM rewriting during execution due to layer-based streaming introduces high latency.

\subsection{Area and Power Breakdown}

%\vspace{-0.3cm}
%\subsection{GEMINI Outperforms State-of-the-art Architectures}
%\vspace{-0.1cm}

%In this section, we compare GEMINI with state-of-the-art reconfigurable dataflow architectures.
%For comparison, we use three typical reconfigurable dataflow architectures: Plasticine \cite{Plasticine}, MTDE \cite{MTDE}, and ParallelXL \cite{ParallelXL}.
%We extend these architectures to have similar peak performance and process technology, using the configurations for clock frequencies from their original papers.
%Due to different design objectives and target applications, there is a significant disparity in peak performance among these designs. We made efforts to mitigate the influence of peak performance and process technology disparity during the evaluation. 
%Table \ref{HC_SOTA} shows the hardware comparisons of our architecture with other designs.

Fig. \ref{A&P} shows the evaluated area and power breakdown of the essential modules in StreamDCIM.
Evaluation shows that StreamDCIM has a chip area of 12.10 mm$^2$ in a 28nm process, and consumes a maximum power of 122.77 mW.
%The highparallelismof the hardware helps toamortize thecontrol overhead.Adetailedsynthesis of the internal structureof PEshows that thehardware for dataflowcontrolon16PEsoccupies3.2\% ofthetotalarea,and thedesignemployed for TimeStamp scheduling incurs little overhead,with0.03\%perPEintotalarea.

\subsection{Performance and Energy Comparison}

\begin{figure}%[htbp]
  \vspace{-0.3cm}
\centering
  \includegraphics[width=0.489\textwidth]{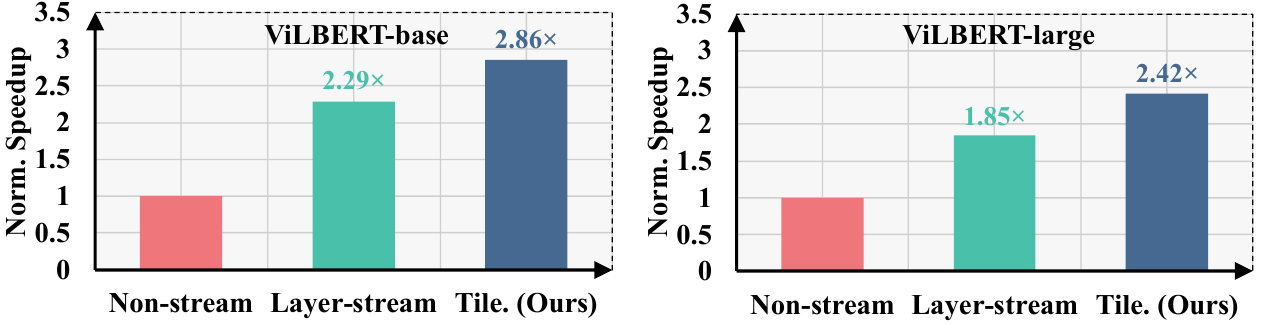}
  \vspace{-0.5cm}
  \caption{Performance Comparison on ViLBERT-base and ViLBERT-large.}
  \label{Speedup}
  \vspace{-0.1cm}
\end{figure}

Fig. \ref{Speedup} illustrates the performance comparison between StreamDCIM and both non-streaming and layer-based streaming solutions on the ViLBERT-base and ViLBERT-large models. 
%On ViLBERT-base, our work achieves 2.86$\times$ and 1.25$\times$ speedup over Non-stream and Layer-stream. On ViLBERT-large, our work achieves 2.42$\times$ and 1.31$\times$ speedup  over Non-stream and Layer-stream. 
Specifically, StreamDCIM achieves a speedup of 2.86$\times$ and 1.25$\times$ over Non-stream and Layer-stream, respectively, on ViLBERT-base model. For ViLBERT-large model, StreamDCIM provides a speedup of 2.42$\times$ compared to Non-stream and 1.31$\times$ over Layer-stream.

\begin{figure}%[htbp]
\vspace{-0.1cm}
\centering
  \includegraphics[width=0.489\textwidth]{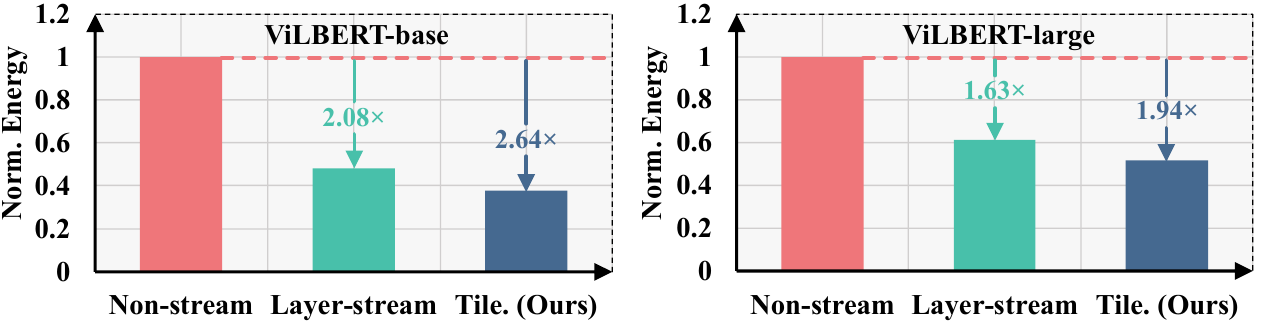}
  \vspace{-0.6cm}
  \caption{Energy Comparison on ViLBERT-base and ViLBERT-large.}
  \label{Energy}
  \vspace{-0.5cm}
\end{figure}

Fig. \ref{Energy} shows the energy comparison, normalized to the non-streaming solution, across the  ViLBERT-base and ViLBERT-large models. 
%On average, our design achieves 1.67$\times$ energy saving over Non-stream and 1.31$\times$ over Layer-stream. 
On ViLBERT-base model, StreamDCIM reduces energy consumption by 2.64$\times$ over Non-stream and 1.27$\times$ over Layer-stream. 
On ViLBERT-large model, StreamDCIM achieves energy saving of 1.94$\times$ compared to Non-stream and 1.19$\times$ over Layer-stream. 
%Plasticine’s static scheduling avoids the runtime overhead of dynamic scheduling and reconfiguration in other dynamic scheduled architectures. However, its lower utilization in most applications results in poor energy efficiency. 
%MTDE’s centralized dynamic scheduling, driven by frequent global load sensing, boosts performance but introduces considerable energy overhead. Despite its performance advantage over ParallelXL, the two have similar energy efficiency. 
%Compared with MTDE, which sacrifices energy for performance, and ParallelXL, which saves energy but also has poorer performance, GEMINI combines on-demand triggering, global sensing \& planning, and low-latency peer-to-peer remapping to maintain high performance without introducing excessive energy overheads, resulting in superior energy efficiency. 

\begin{comment}
%\vspace{-0.2cm}
\subsection{Overhead}
%\subsection{Cost and Discussions}
%\vspace{-0.1cm}

Finally, we evaluate the hardware overhead of the adaptive prefetching and decentralized scheduling. First is the adaptive prefetching. The hardware overhead of PANDA MEM is shown in Table \ref{SPR}, and PANDA MEM exhibits a total area overhead of 37.4\% and 12.9\% increase over SPM with DMA and cache of the same size, respectively. 
Next is the decentralized scheduling. The hardware overhead of PANDA PE is shown in Table \ref{CD}, and PANDA PE exhibits a total area overhead of 6.5\% increase over the Base PE of the same computing fabric. 
\end{comment}

\section{Conclusion}

%In this paper, we describe GEMINI, a novel dataflow architecture with decoupled task flow and data flow planes, dedicated to efficient dynamic scheduling through hardware-software co-optimizatios. 
In this paper, we describe StreamDCIM, a tile-based streaming digital CIM accelerator with mixed-stationary cross-forwarding dataflow for multimodal Transformers.
%We implement GEMINI using Verilog and demonstrate that in a wide range of real-world applications, including scientific computing, artificial intelligence, signal processing, and graph processing algorithms, GEMINI attains up to 1.89$\times$ performance and 1.67$\times$ energy efficiency improvement over the state-of-the-art dataflow architectures.
We implement StreamDCIM using Verilog and demonstrate that comparede with non-streaming and layer-based streaming CIM-based solutions, StreamDCIM attains by geomean 2.63$\times$, 1.28$\times$ speedup and 2.26$\times$, 1.23$\times$ energy saving on typical multimodal Transformer models.%, including scientific computing, artificial intelligence, signal processing, and graph processing algorithms.

%\begin{comment}
\section*{Acknowledgment}

This work was supported by Beijing Natural Science Foundation (Grant No.L234078) and Beijing Nova Program (Grant No. 20220484054 and No. 20230484420).
%\end{comment}

\vspace{12pt}
%\color{red}
%IEEE conference templates contain guidance text for composing and formatting conference papers. Please ensure that all template text is removed from your conference paper prior to submission to the conference. Failure to remove the template text from your paper may result in your paper not being published.

\end{document}